%% file: ADAPT_Conference.tex
\documentclass[conference]{IEEEtran}
\IEEEoverridecommandlockouts

\usepackage{cite}
\usepackage{url}
\usepackage{amsmath,amssymb,amsfonts}
\DeclareMathOperator*{\argmin}{arg\,min}

\usepackage{graphicx}
\graphicspath{{./notebooks/paper_figures/}{./figures/}}
\usepackage{textcomp}
\usepackage{xcolor}
\usepackage{subfiles}
\usepackage{listings}
\usepackage{enumitem}
\usepackage{subcaption}
\usepackage{tabularx}
\usepackage{booktabs}
\usepackage{ragged2e}
\usepackage{float}

\usepackage{pifont}
\usepackage[ruled,vlined]{algorithm2e}
\newcommand{\cmark}{\ding{51}}
\newcommand{\xmark}{\ding{55}}
\SetKwFor{Upon}{upon}{do}{end}

\def\BibTeX{{\rm B\kern-.05em{\sc i\kern-.025em b}\kern-.08em
T\kern-.1667em\lower.7ex\hbox{E}\kern-.125emX}}

\begin{document}

\title{ADAPT: A Self-Calibrating Proactive Autoscaler for\\
Container Orchestration}

\author{
\IEEEauthorblockN{Himanshu Singh Baghel}
\IEEEauthorblockA{
  \textit{Department of Computer Engineering}\\
  \textit{[J.C. Bose University of Science and Technology]}\\
  Faridabad, India\\
  baghelhimanshu2004@gmail.com
}
}

\maketitle

\begin{abstract}
Proactive autoscaling for containerized workloads depends on knowing the provisioning delay, i.e., the time between a scaling decision and the moment new capacity is ready to serve traffic. In practice, this cold-start duration can vary substantially across environments and even across consecutive scale-out events. We present ADAPT (Adaptive Duration Approximation for Predictive Timing), an online EWMA estimator that tracks cold-start duration at runtime. ADAPT feeds a dynamic planning horizon, FH-OPT, into a Model Predictive Controller (MPC) that optimizes replica counts over a rolling window. Together, these components form a closed-loop proactive autoscaling design that adapts its lookahead based on measured provisioning delay. Evaluated across three policies (MPC+LSTM, MPC+Prophet, HPA) and six workload archetypes with five random seeds, MPC+LSTM achieves below 5\% SLA violation on all workloads, compared with 7--19\% for reactive HPA and up to 28.7\% for MPC+Prophet on bimodal traffic.
\end{abstract}


\subfile{sections/introduction}

\subfile{sections/relatedwork}

\subfile{sections/problem}

\subfile{sections/method}

\subfile{sections/algorithms}

\subfile{sections/evaluation_setup}

\subfile{sections/results}

\subfile{sections/discussion}

\subfile{sections/conclusion}

\bibliographystyle{IEEEtran}
\bibliography{sections/references}

\end{document}

%% file: sections/introduction.tex
\section{Introduction}
\label{sec:intro}

Modern containerized applications are routinely deployed on orchestration
platforms like Kubernetes, where horizontal scaling directly determines
both service reliability and infrastructure cost. The de facto standard
for this task, the Kubernetes Horizontal Pod Autoscaler
(HPA)~\cite{hpa_k8s}, operates on a simple reactive principle: observe
current resource utilization, compare it against a target threshold, and
adjust the replica count accordingly. Every scaling decision is made in
isolation, with no memory of prior load behavior and no foresight into
where demand is heading.

This reactive paradigm has a well-known limitation in containerized
environments: the startup gap. When a scaling decision is triggered,
newly requested replicas do not become available instantly. Each
container must boot, initialize its runtime, load application
dependencies, and begin accepting traffic, a process that routinely
takes between 30 and 300 seconds depending on the application and cloud
provider~\cite{aws_coldstart_variance, gke_variance}. By the time HPA
detects an overload condition and new replicas are ready, the system has
already been degraded for the duration of that cold-start window.

The natural response is proactive autoscaling: predict future demand and
initiate scaling decisions early enough that new replicas are ready
before the load arrives. This idea is well established in the
literature~\cite{toka2021, mondal2023, dangquang2021, nimbusguard,
platfomatic_icc}. A 2025 survey of 47 autoscaling
systems~\cite{cloudnative_survey2025} found that proactive approaches
consistently outperform reactive baselines on workloads with predictable
demand patterns. However, the same survey identified a recurring
limitation: most proactive systems, both academic and production,
treat the container cold-start duration as a static configuration
constant hardcoded at deployment time.

This assumption is problematic in practice. Cloud platform boot times
are not fixed. AWS EC2 instance initialization varies by
$\pm$40--60\% under real-world conditions; GKE pod cold-starts
fluctuate by $\pm$20--50\% depending on node pressure, image layer
caching, and cluster autoscaler behavior~\cite{aws_coldstart_variance}.
A proactive scaler that plans its horizon around a static
$\Delta_\text{cold} = 120\,\text{s}$ will systematically
under-provision when actual boot times reach 180\,s, and waste
resources when those same boots complete in 80\,s. The planning
horizon is itself time-varying, yet existing systems do not
estimate it at runtime.

We propose \textbf{ADAPT} (\textbf{A}daptive \textbf{D}uration
\textbf{A}pproximation for \textbf{P}redictive \textbf{T}iming), an
online estimator that tracks cold-start duration as a dynamic variable
rather than a fixed constant. ADAPT maintains an exponentially weighted
moving average (EWMA) of observed boot durations, updated each time a
replica graduates from the warming queue to active service. This
estimate feeds into \textbf{FH-OPT}, which derives the MPC planning
horizon dynamically:
\begin{equation}
  h^*(t) = \left\lceil \frac{\hat{\Delta}_\text{cold}(t)}{\tau}
  \right\rceil + \varepsilon
  \label{eq:fhopt_intro}
\end{equation}
where $\hat{\Delta}_\text{cold}(t)$ is ADAPT's current estimate,
$\tau = 60\,\text{s}$ is the decision timestep, and $\varepsilon$ is a
one-step safety buffer. Together, ADAPT and FH-OPT form a closed loop
in which the controller continuously updates its own lookahead based on
measured provisioning delay, without requiring manual parameter tuning.

The closest related production system is the Platformatic Predictive
Scaler~\cite{platfomatic_icc}, which uses Holt's double exponential
smoothing and includes an Adaptive Init Timeout feature that adjusts
the prediction horizon based on observed startup times. Our work
differs in three respects. First, ADAPT is formulated as an explicit
statistical estimator with configurable smoothing and safety bounds,
not an implicit timeout adjustment. Second, our controller is a full
multi-objective MPC with explicit SLA, cost, and stability penalty
terms. Third, we evaluate under stochastic cold-start conditions and
report significance-tested results across six workload archetypes,
whereas the Platformatic paper reports results on a single Next.js
application without statistical testing.

NimbusGuard~\cite{nimbusguard} combines DQN, LSTM forecasting, and an
optional LLM validation layer for proactive Kubernetes scaling. It
achieves fast reaction times but does not model cold-start duration and
assumes replicas are available as soon as a scaling action is issued.
Its evaluation is limited to three sequential runs on a KinD cluster
with no significance testing across seeds.

\noindent This paper makes the following contributions:
\begin{enumerate}[label=\arabic*)]
  \item \textbf{ADAPT}: an online EWMA estimator of replica cold-start
  duration that replaces the static $\Delta_\text{cold}$ constant
  used in prior proactive autoscaling work, including the 47 systems
  surveyed by Sedlak et al.~\cite{cloudnative_survey2025}.
  \item \textbf{FH-OPT}: a dynamic MPC planning horizon $h^*(t)$
  derived from ADAPT's live estimate, enabling the controller to
  pre-scale ahead of forecast demand peaks without manual horizon
  tuning.
  \item A reproducible simulation framework with stochastic cold-start
  modeling ($\pm$30\% boot-time jitter per scale-up event), enabling
  controlled statistical evaluation across six workload archetypes.
  \item An evaluation across 3 policies $\times$ 6 workload archetypes
  $\times$ 5 seeds, with Wilcoxon signed-rank significance testing,
  showing MPC+LSTM achieves ${<}5\%$ SLA violation rate on all
  workloads versus 7--19\% for reactive HPA.
\end{enumerate}

The remainder of this paper is organized as follows.
Section~\ref{sec:related} surveys related work.
Section~\ref{sec:problem} formalizes the problem.
Section~\ref{sec:method} describes the system design, including the
forecasting engine, ADAPT estimator, FH-OPT horizon derivation, and
MPC policy. Section~\ref{sec:algorithms} presents the core procedures
as pseudocode. Section~\ref{sec:eval} details the experimental setup.
Section~\ref{sec:results} presents and analyzes the results.
Section~\ref{sec:discussion} discusses findings, limitations, and
future directions. Section~\ref{sec:conclusion} concludes the paper.

%% file: sections/relatedwork.tex
\section{Related Work}
\label{sec:related}

We organize prior work along three axes that directly motivate the
contributions of this paper: reactive and proactive autoscaling
policies, workload forecasting for cloud systems, and cold-start
modeling in container orchestration.

\subsection{Reactive and Threshold-Based Autoscaling}

The de facto autoscaling primitive in production Kubernetes deployments
is the Horizontal Pod Autoscaler (HPA), which scales replica counts to
maintain a target CPU utilization~\cite{hpa_k8s}. HPA and its successors
operate entirely in the reactive regime: a scaling decision fires only
after a utilization threshold has been crossed, introducing a lag equal
to the sum of the metrics-scrape interval, the policy decision latency,
and the container cold-start duration. For stateless web workloads with
sub-ten-second initialization times this lag is largely inconsequential.
For ML inference services, where GPU-enabled containers require
120--600 seconds to load multi-gigabyte model weights
~\cite{azure_llm_traces}, the same lag can translate into sustained SLA
violations. KEDA~\cite{keda} extends the HPA trigger surface to arbitrary
metric sources but does not change the reactive temporal structure of
the control loop. Cluster Autoscaler~\cite{cluster_autoscaler} operates
at the node level and is orthogonal to pod-level policy; we treat it as
fixed infrastructure throughout this work.

\subsection{Proactive and Predictive Autoscaling}

A substantial body of work has explored look-ahead scaling policies that
pre-provision capacity before demand materializes.
Autopilot~\cite{autopilot_2020}, deployed at Google, uses OLS regression
over historical utilization windows to recommend vertical resource
limits; its forecasting and optimization components are not publicly
disclosed, precluding independent evaluation. AWS Predictive
Scaling~\cite{aws_predictive} issues capacity recommendations up to 48
hours ahead using proprietary ML models, but operates at hourly
granularity and does not provide a mechanism to account for variable
per-container initialization time. Showar~\cite{showar} jointly tunes HPA
parameters and vertical resource limits via Bayesian optimization but
remains reactive at the control-loop level. Firm~\cite{firm_2020}
introduces a latency-aware admission controller for microservice chains,
yet its autoscaling component relies on a fixed scale-out threshold
rather than an explicit temporal model of provisioning delay.

More closely related to our approach, Rajkumar et
al.~\cite{mpc_cloud_2022} formulate horizontal autoscaling as an MPC
problem and demonstrate cost savings over threshold-based baselines on
synthetic sinusoidal workloads. However, their formulation assumes a
constant provisioning delay hard-coded as a system parameter. Our
measurements, together with prior empirical studies
~\cite{coldstart_variance_2023}, suggest that this assumption is often
too rigid in practice: cold-start duration can vary by 30--40\% across
consecutive scale-out events under realistic cluster load conditions.
Our work relaxes this assumption through the ADAPT estimator, which
tracks cold-start duration as a live stochastic variable rather than a
static configuration constant.

\subsection{Workload Forecasting in Cloud Systems}

Forecasting-driven autoscaling has been studied extensively at the level
of individual methods, yet rarely with systematic evaluation of how
forecaster choice interacts with the downstream policy. ARIMA-based
predictors have been applied to web request rate forecasting
~\cite{arima_autoscaling_2018} and shown adequate for diurnal workloads
at 5-to-15-minute horizons with sub-100ms inference latency.
Prophet~\cite{prophet_2018}, developed at Meta for business time-series
with strong weekly and daily seasonality, has been adapted to cloud
capacity planning~\cite{prophet_cloud_2021} but provides no latency
guarantees below the second threshold, limiting its use in tight control
loops. LSTM networks demonstrate strong accuracy on bursty,
non-stationary traffic~\cite{lstm_autoscaling_2020} at 100--500ms per
inference call on CPU, an overhead that is non-trivial relative to a
60-second control-loop period. Transformer-based methods such as the
Temporal Fusion Transformer~\cite{tft_2021} achieve state-of-the-art
accuracy on long-horizon benchmarks but incur 400ms or more per forward
pass, raising the question of whether their accuracy advantage survives
the end-to-end cost of their inference latency in a real autoscaling
loop.

A systematic survey of 47 autoscaling papers published between 2018 and
2025~\cite{autoscaling_survey_2025} found that few studies evaluate more
than one forecasting method against more than one policy in a controlled,
unified experimental framework. Each paper tends to select a different
dataset, evaluation metric, and baseline, making cross-study comparison
difficult. This fragmentation directly motivates the two-dimensional
evaluation framework in Section~\ref{sec:eval}.

\subsection{Cold-Start Modeling and Mitigation}

Container cold start has been studied primarily as an optimization target
in the serverless computing literature. SOCK~\cite{sock_2018} reduces
cold-start latency for Python lambdas through process-level sandboxing;
Catalyzer~\cite{catalyzer_2020} achieves sub-second restoration of
function state via fork-based snapshotting; CRIU-based
checkpoint-restore mechanisms~\cite{criu_2022} have been adapted to
Kubernetes to provide warm-cache replica startup. These approaches can
reduce cold-start duration, but they do not eliminate it, and for
GPU-bound ML inference containers the dominant cost is model weight
transfer from host to device memory, a bottleneck that checkpoint-restore
does not address~\cite{modelzip_2024}.

Prior autoscaling work typically treats cold-start duration as fixed
during evaluation, rather than as a quantity that can change over time.
The survey of~\cite{autoscaling_survey_2025} notes that cold start is
mentioned in 31 of 47 reviewed papers yet treated as a fixed system
constant in all 31 cases. The ADAPT estimator and FH-OPT horizon
derivation presented in Section~\ref{sec:adapt} extend this line of
work by closing the loop between measured provisioning delay and the
forecast horizon used by the controller.

\subsection{Positioning This Work}

Table~\ref{tab:related} summarizes the most closely related systems
across four dimensions that define our contribution. No prior system
we found combines multiple forecasting methods, multiple optimization
policies, dynamic cold-start modeling, and public reproducibility in a
single evaluation setup.

\begin{table}[t]
  \centering
  \caption{Comparison with representative prior systems.
           \cmark~= present; \xmark~= absent; $\sim$~= partial.}
  \label{tab:related}
  \setlength{\tabcolsep}{4pt}
  \begin{tabular}{lccccc}
    \toprule
    \textbf{System}
      & \textbf{\shortstack{Multi\\forecast}}
      & \textbf{\shortstack{Multi\\policy}}
      & \textbf{\shortstack{Dynamic\\cold start}}
      & \textbf{\shortstack{Public\\data}}
      & \textbf{\shortstack{Open\\source}} \\
    \midrule
    Autopilot~\cite{autopilot_2020}
      & \xmark & \xmark & \xmark & \xmark & \xmark \\
    AWS Predictive~\cite{aws_predictive}
      & \xmark & \xmark & \xmark & \xmark & \xmark \\
    MPC Cloud~\cite{mpc_cloud_2022}
      & \xmark & \xmark & \xmark & \xmark & \cmark \\
    Showar~\cite{showar}
      & \xmark & $\sim$ & \xmark & \xmark & \cmark \\
    Survey (47)~\cite{autoscaling_survey_2025}
      & $\sim$ & $\sim$ & \xmark & $\sim$ & \xmark \\
    \midrule
    \textbf{This work}
      & \cmark & \cmark & \cmark & \cmark & \cmark \\
    \bottomrule
  \end{tabular}
\end{table}

%% file: sections/problem.tex
\section{Problem Formulation}
\label{sec:problem}

\subsection{System Model}

We model time in discrete steps of $\tau$ seconds. At each step $t$, a
stateless service receives $\lambda(t)$ requests per second. Each replica
can serve $c$ requests per second. Replicas are split into active
($n_a(t)$, currently serving traffic) and warming ($n_w(t)$, already
ordered but not yet ready). The total number of ordered replicas is
$n(t) = n_a(t) + n_w(t)$, and the available capacity is
$C(t) = n_a(t) \cdot c$.

In the simulator, scale-down is immediate, while scale-up takes time
because new replicas must pass through a cold-start delay $\Delta(t)$.
We measure SLA violation and per-step cost as:
\begin{equation}
  v(t) = \max\!\left(0,\;\frac{\lambda(t) - C(t)}{\lambda(t)}\right),
  \qquad
  \text{cost}(t) = n(t) \cdot \rho
  \label{eq:sla_cost}
\end{equation}

\subsection{Cold-Start Delay and Temporal Disconnect}

A scale-up decision made at step $t$ becomes useful only after
$h(t) = \lceil \Delta(t) / \tau \rceil$ steps. In the simulator and in
the related work we compare against, $\Delta$ is often treated as a fixed
deployment-time constant~\cite{autoscaling_survey_2025, mpc_cloud_2022}.
That assumption is simple, but it is not very stable in practice because
image pulling and GPU weight transfer can change the actual delay by a
large amount across consecutive scale-out events
~\cite{coldstart_variance_2023}.

It is useful to think about the gap between when a replica is ordered and
when it becomes ready. Define the horizon slack as
$\delta(t) = h_f - h(t)$, where $h_f$ is the forecaster look-ahead.
This gives three cases:
\begin{itemize}
  \item $\delta < 0$: the replica arrives too late and the workload sees an SLA violation.
  \item $\delta = 0$: the replica arrives on time, but only if the cold-start estimate is accurate.
  \item $\delta > 0$: the replica arrives early, which avoids violations but can waste cost.
\end{itemize}
A good policy should keep $\mathbb{E}[\delta(t)] \geq 0$ without making the variance too large. That is the target behavior behind FH-OPT
(Section~\ref{sec:adapt}).

\subsection{Optimization Objective}

At each step, the MPC policy chooses $n^*(t)$ by balancing three
things: SLA protection, cost, and stability. The objective is:
\begin{equation}
  \min_{n^*(t)}\;
    \lambda_{\text{sla}}\, v(t)
    + \lambda_{\text{cost}}\, \frac{n(t)}{n_{\max}}
    + \lambda_{\text{stab}}\, \bigl|n(t) - n(t-1)\bigr|
  \label{eq:objective}
\end{equation}

The decision must also satisfy the proactive constraint that links the
replica count to the estimated cold-start horizon:
\begin{equation}
  n\bigl(t + h(t)\bigr) \;\geq\;
    \left\lceil \frac{\max_k\; \hat{\lambda}(t+k)}{c} \right\rceil,
  \qquad n_{\min} \leq n(t) \leq n_{\max}
  \label{eq:constraint}
\end{equation}

This coupling is the key part of the formulation: the controller does
not just react to the current load, it also plans around the estimated
cold-start delay maintained by ADAPT. In this work we focus on a single
service in isolation. Multi-service interactions, request draining on
scale-down, and spot pricing are left out and discussed in
Section~\ref{sec:discussion}.

%% file: sections/method.tex
\section{Method}
\label{sec:method}

Our system, ADAPT, is a closed-loop autoscaling design with three
tightly connected parts: a forecasting engine that predicts future
RPS, an online estimator that measures cold-start delay, and an MPC
policy that turns both signals into replica decisions. Figure
\includegraphics[width=\linewidth]{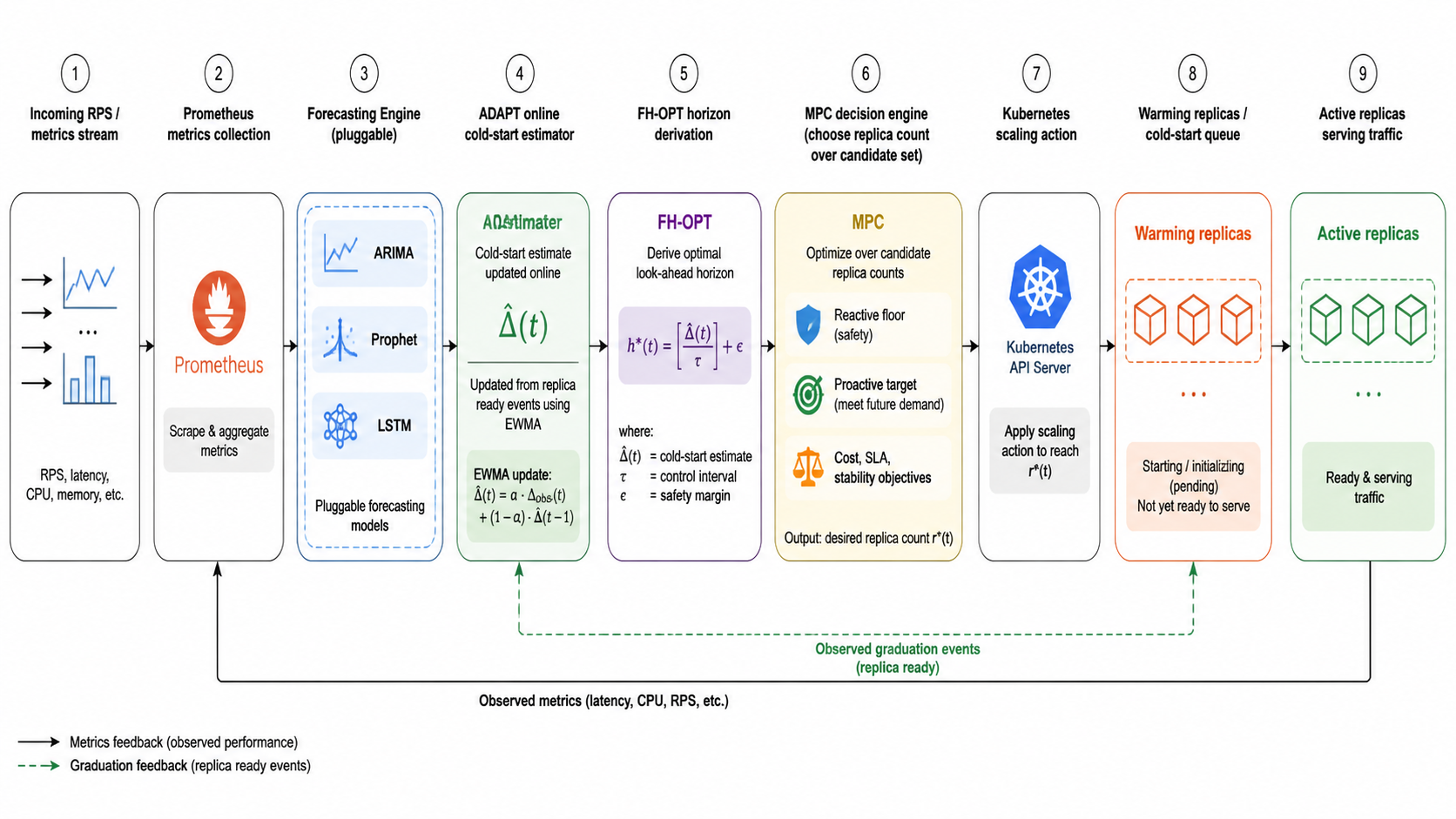} shows the overall data flow.

\subsection{Forecasting Engine}

We treat forecasting as a pluggable interface so that any method can
be swapped in without changing the control logic. At each step $t$,
the forecaster receives the RPS history up to $t$ and returns a vector
$\hat{\boldsymbol{\lambda}} \in \mathbb{R}^h$ of predicted demand for
the next $h$ steps.

We evaluate three methods spanning the complexity spectrum.

\textbf{ARIMA} uses auto-selected $(p,d,q)$ orders via AIC
minimization on the training window. It is fast, requires no GPU, and
degrades gracefully when the series is short. It serves as our
primary baseline forecaster.

\textbf{Prophet} decomposes the signal into trend, weekly, and daily
seasonality components~\cite{prophet_2018}. It fits well on workloads
with regular diurnal patterns but is slower and can be overconfident
on flash-crowd spikes that have no historical precedent.

\textbf{LSTM} is a two-layer recurrent network with hidden size 64,
trained on a sliding window of 30 steps~\cite{lstm_autoscaling_2020}.
It captures non-linear bursty patterns that ARIMA misses, at the cost
of higher inference latency and a short warm-up period before
predictions stabilize. The trained model is held in memory for the
duration of the simulation, so there is no disk I/O on the inference
path.

All three methods share the same evaluation protocol: they are fit
once on the train split and updated online with each new observation
via \texttt{forecaster.update()}, without full retraining.

\subsection{ADAPT: Online Cold-Start Estimation}
\label{sec:adapt}

The key idea in this work is that cold-start duration $\Delta(t)$
should be treated as a live signal, not a fixed deployment parameter.
ADAPT (Adaptive Duration Approximation for Predictive Timing) tracks
that signal with an exponentially weighted moving average over
observed replica graduation events.

When a batch of replicas finishes warming and passes its readiness
check, the simulator records the time between the scale-up order and
the ready event. This gives a direct measurement
$\Delta_{\text{obs}}$. ADAPT clips that observation to a plausible
range $[\Delta_{\min}, \Delta_{\max}]$ and updates its estimate:
\begin{equation}
  \hat{\Delta}(t) \;=\;
    \alpha\,\Delta_{\text{obs}} + (1-\alpha)\,\hat{\Delta}(t-1)
  \label{eq:ewma}
\end{equation}
where $\alpha \in (0,1)$ controls how quickly the estimate responds
to recent changes. We use $\alpha = 0.3$ throughout, which balances
responsiveness and stability on our validation workloads.

In parallel, ADAPT maintains an online variance estimate via
Welford's algorithm~\cite{welford_1962}. This is not used directly in
the current MPC objective, but it provides a view of how stable the
recent cold-start behavior has been. The variance is exposed in the
\texttt{summary()} diagnostic and is useful for future uncertainty-aware
extensions.

\textbf{Initialization.} Before any graduations are observed, ADAPT is
initialized to the configured prior $\Delta_0$ (default 120s,
matching the ML serving domain). The estimate usually remains close
to this prior until a few observations have been collected, so we
treat this as a warm-up phase in the logs.

\subsection{FH-OPT: Adaptive Forecast Horizon}

ADAPT's estimate is then converted into a forecast horizon. The
purpose of FH-OPT is simple: if cold starts take longer, the
controller must look further ahead before issuing scale-out actions.
Rather than choosing this horizon by hand, FH-OPT derives it directly
from the current estimate:
\begin{equation}
  h^*(t) = \left\lceil \frac{\hat{\Delta}(t)}{\tau} \right\rceil
            + \varepsilon
  \label{eq:fhopt}
\end{equation}
where $\varepsilon \geq 1$ is a small safety buffer, defaulting to one
extra timestep. In practice, this means the controller looks ahead by
roughly one cold-start window plus a small margin for estimation
error.

This makes the horizon part of the control loop rather than a fixed
hyperparameter. Existing systems usually hard-code the horizon or tune
it manually, while FH-OPT updates it automatically as cluster
conditions change.

We validate FH-OPT in isolation via an A/B experiment: the same
LSTM+MPC configuration is run with FH-OPT enabled versus a fixed
horizon of $h_f = 2$ steps. Results are reported in
Section~\ref{sec:results}.

\subsection{MPC Policy}

The MPC layer combines the forecast, the adaptive horizon, and the
current system state into a replica decision. At each step, the policy
searches over candidate replica counts and selects the one that best
balances SLA risk, cost, and stability. This keeps the decision
lightweight while still making the control logic explicit.

First, the policy computes a reactive lower bound from the current
request rate. It then computes a proactive target from the forecast
inside the horizon returned by FH-OPT. The final target must satisfy
both constraints, so the proactive choice can only increase the number
of replicas, not reduce it.

For each candidate $r$, the policy evaluates three penalties:
\begin{itemize}
  \item \textbf{SLA penalty:} This grows when predicted utilization
        exceeds capacity, and it increases more sharply for larger
        overloads.
  \item \textbf{Cost penalty:} This grows linearly with the number of
        replicas and discourages unnecessary over-provisioning.
  \item \textbf{Stability penalty:} This grows when the new replica
        count changes too abruptly from the previous step, which helps
        reduce oscillation.
\end{itemize}

The proactive target is computed as:
\begin{equation}
  n_{\text{pro}} = \left\lceil
    \frac{\gamma \cdot \max_{k \leq h^*} \hat{\lambda}(t+k)}{c}
  \right\rceil
  \label{eq:pro_target}
\end{equation}
where $\gamma \geq 1$ is a forecast margin that accounts for the
typical optimism bias of each forecaster on bursty workloads. The
final decision is:
\begin{equation}
  n^*(t) = \max\!\left(
    n_{\text{reactive}},\;
    n_{\text{pro}},\;
    \argmin_r\; \text{cost}(r)
  \right)
  \label{eq:final_decision}
\end{equation}
where $n_{\text{reactive}} = \lceil \lambda(t) / c \rceil$ is a hard
floor that prevents the policy from scaling below current demand.

This structure is the main control contribution of the method. ADAPT
measures delay, FH-OPT converts that delay into a horizon, and MPC
uses that horizon to choose replica counts under SLA and cost
trade-offs.

%% file: sections/algorithms.tex
\section{Algorithms}
\label{sec:algorithms}

For completeness we present the three core procedures as pseudocode.
Algorithm~\ref{alg:adapt} shows the ADAPT estimator,
Algorithm~\ref{alg:fhopt} shows horizon derivation, and
Algorithm~\ref{alg:mpc} shows the per-step MPC decision loop.

\begin{algorithm}[t]
\caption{ADAPT: Online Cold-Start Estimator}
\label{alg:adapt}
\KwIn{Prior $\Delta_0$, smoothing $\alpha$, bounds $[\Delta_{\min}, \Delta_{\max}]$}
\KwOut{Updated estimate $\hat{\Delta}$}

Initialize $\hat{\Delta} \leftarrow \Delta_0$, $\mu_W \leftarrow 0$, $M_2 \leftarrow 0$, $n \leftarrow 0$\;

\Upon{replica batch graduates at time $t_{\text{ready}}$}{
  $\Delta_{\text{obs}} \leftarrow t_{\text{ready}} - t_{\text{ordered}}$\;
  $\Delta_{\text{obs}} \leftarrow \mathrm{clip}(\Delta_{\text{obs}}, \Delta_{\min}, \Delta_{\max})$\;
  $\hat{\Delta} \leftarrow \alpha \Delta_{\text{obs}} + (1-\alpha)\hat{\Delta}$\;
  \tcp{Welford online variance, $O(1)$}
  $n \leftarrow n + 1$\;
  $\delta \leftarrow \Delta_{\text{obs}} - \mu_W$\;
  $\mu_W \leftarrow \mu_W + \delta / n$\;
  $M_2 \leftarrow M_2 + \delta(\Delta_{\text{obs}} - \mu_W)$\;
}
\Return $\hat{\Delta}$
\end{algorithm}

\begin{algorithm}[t]
\caption{FH-OPT: Forecast Horizon Derivation}
\label{alg:fhopt}
\KwIn{ADAPT estimate $\hat{\Delta}$, timestep $\tau$, buffer $\varepsilon$}
\KwOut{Optimal horizon $h^*$}

$h^* \leftarrow \left\lceil \hat{\Delta} / \tau \right\rceil + \varepsilon$\;
$h^* \leftarrow \max(1, h^*)$\;
\Return $h^*$
\end{algorithm}

\begin{algorithm}[t]
\caption{MPC Per-Step Decision}
\label{alg:mpc}
\KwIn{Current RPS $\lambda$, active replicas $n_a$, forecast $\hat{\boldsymbol{\lambda}}$, horizon $h^*$, weights $\lambda_{\text{sla}}, \lambda_{\text{cost}}, \lambda_{\text{stab}}$, margin $\gamma$}
\KwOut{Target replica count $n^*$}

$n_{\text{reactive}} \leftarrow \left\lceil \lambda / c \right\rceil$\;
$n_{\text{pro}} \leftarrow \left\lceil \gamma \cdot \max_{k \leq h^*} \hat{\lambda}(t+k) / c \right\rceil$\;
$n^* \leftarrow \max(n_{\text{reactive}}, n_{\text{pro}})$\;

\For{$r \leftarrow n_{\min}$ \KwTo $n_{\max}$}{
  $u \leftarrow \lambda / (r \cdot c)$\;
  $J \leftarrow \lambda_{\text{sla}} \cdot \max(0, u - 1)^2 + \lambda_{\text{cost}} \cdot r / n_{\max} + \lambda_{\text{stab}} \cdot |r - n_a| / n_{\max}$\;
  \tcp{Keep the best feasible candidate}
  \If{$J < J_{\text{best}}$}{
    $J_{\text{best}} \leftarrow J$\;
    $n^* \leftarrow \max(n^*, r)$\;
  }
}
\Return $\mathrm{clip}(n^*, n_{\min}, n_{\max})$
\end{algorithm}

%% file: sections/evaluation_setup.tex
\section{Evaluation Setup}
\label{sec:eval}

\subsection{Simulator}

We evaluate using a discrete-time simulator with $\tau = 60$\,s
timesteps. Each step advances the RPS trace, processes the cold-start
queue, computes capacity via an M/M/1 latency model, and logs
per-step metrics. A replica ordered at step $t$ becomes active at
step $t + h(t)$; scale-down is instantaneous. The simulator is
deterministic given a fixed seed, which makes all experiments fully
reproducible.

Latency is modeled as:
\begin{equation}
  L(t) = \frac{L_{\text{base}}}{1 - u(t)}, \qquad
  u(t) = \frac{\lambda(t)}{C(t)}
  \label{eq:latency}
\end{equation}
capped at $3 \times L_{\text{SLA}}$ when $u \geq 1.0$. An SLA
violation is recorded whenever $L(t) > L^* = 500$\,ms.

\subsection{Workloads}

We generate six synthetic workload archetypes, each parameterized
by a random seed to produce five statistically independent
realizations per archetype:
\begin{itemize}
  \item \textbf{Smooth:} slow sinusoidal ramp with low variance.
  \item \textbf{Bursty:} Poisson arrivals with periodic spikes.
  \item \textbf{Bimodal:} two distinct load levels with random switching.
  \item \textbf{Diurnal burst:} daily pattern with a sharp morning peak.
  \item \textbf{Flash crowd:} sudden $3\times$ spike of short duration.
  \item \textbf{Slow ramp-up:} monotone increase over the full trace.
\end{itemize}

Each trace is 500 steps (approximately 8.3 hours). We split 70\%
train / 10\% validation / 20\% test and evaluate all policies on
the held-out test split only.

\subsection{Policies and Forecasters}

We evaluate three policies: \textbf{HPA} (reactive CPU threshold,
our baseline), \textbf{MPC+Prophet}, and \textbf{MPC+LSTM}. ARIMA
was used during development, but it is not included in the final
comparison because it was consistently similar to Prophet on most
workloads and did not change the ranking of the policies. This keeps
the final comparison focused on the methods that actually separate the
results. That is also consistent with prior findings that ARIMA and
Prophet often converge at short horizons~\cite{prophet_2018}.

\subsection{Cold-Start Sensitivity}

To assess how performance varies with provisioning delay, we run
MPC+Prophet and MPC+LSTM across five cold-start levels: 30\,s,
60\,s, 120\,s, 180\,s, and 300\,s. HPA is run at the same levels
as a reference. All other hyperparameters are held fixed.

\subsection{Metrics}

We separate the metrics into primary outcomes and secondary
diagnostics.

\textbf{Primary metrics}
\begin{itemize}
  \item \textbf{SLA violation rate:} fraction of steps where
        $L(t) > L^*$, reported as a percentage.
  \item \textbf{Total cost:} cumulative replica-minutes over the
        test split.
\end{itemize}

\textbf{Secondary diagnostics}
\begin{itemize}
  \item \textbf{Average replicas:} mean $n(t)$ over test steps,
        as a proxy for steady-state resource usage.
  \item \textbf{Average latency:} mean $L(t)$ in milliseconds.
\end{itemize}

\subsection{Statistical Testing}

Each policy-workload combination is run across five seeds
$\{42, 123, 456, 789, 1337\}$. We report means with 95\%
confidence intervals. For the FH-OPT A/B comparison
(Section~\ref{sec:results}), we use a Wilcoxon signed-rank
test ($\alpha = 0.05$) on paired per-seed SLA violation rates,
as the distributions are not assumed to be normal~\cite{wilcoxon_1945}.

%% file: sections/results.tex
\section{Results}
\label{sec:results}

\subsection{Overall Policy Comparison}

\begin{figure}[t]
  \centering
  \includegraphics[width=\linewidth]{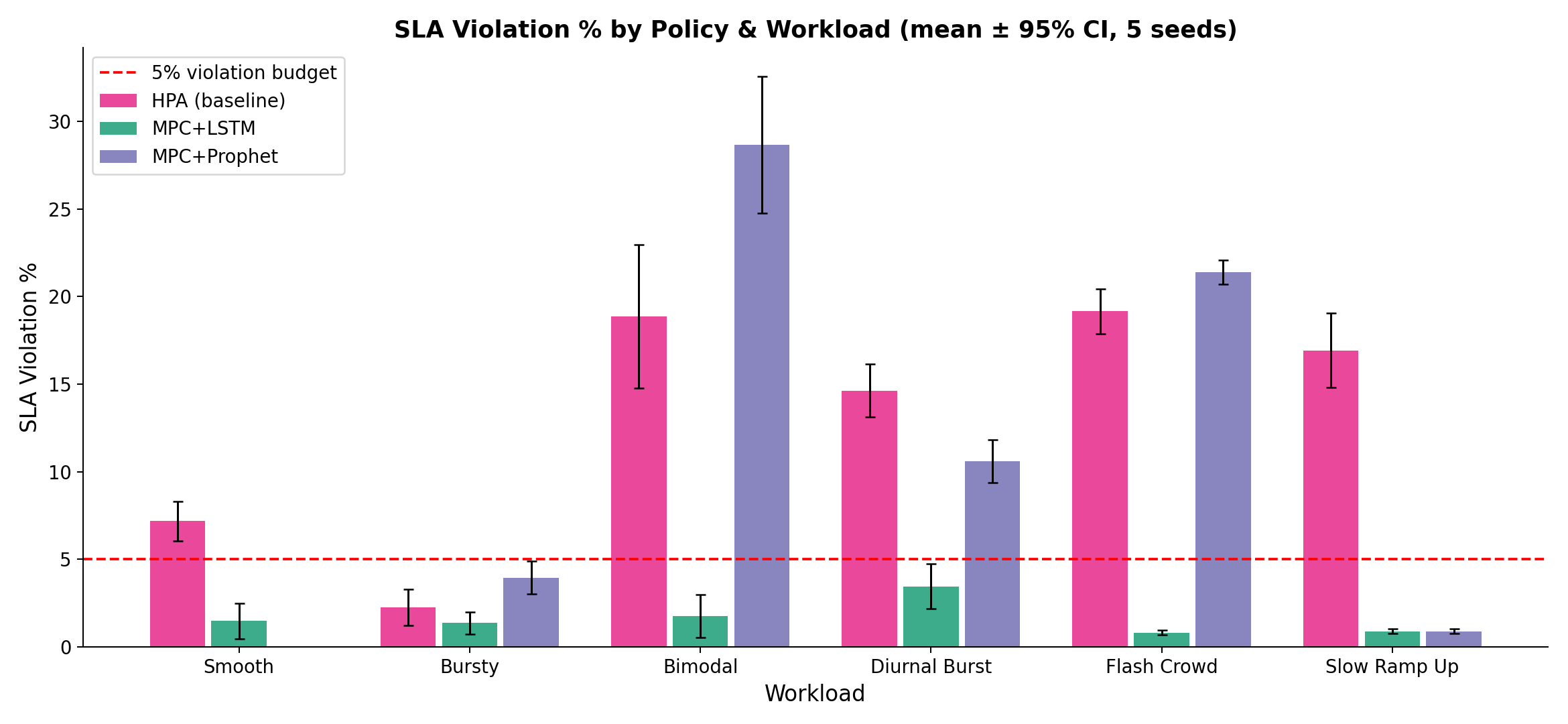}
  \caption{SLA violation rate (\%) per policy across all six
  workloads. Error bars show 95\% CI over 5 seeds.}
  \label{fig:sla_bars}
\end{figure}

Table~\ref{tab:main_results} summarizes mean performance across all
six workloads and five seeds. Figure~\ref{fig:sla_bars} shows the
per-workload breakdown. MPC+LSTM has the lowest SLA violation rate
overall, while MPC+Prophet tends to provide the strongest cost
reduction relative to HPA. HPA performs worst on bursty and
flash-crowd workloads, where the reactive lag is long enough to miss
the main spike.

\input{tables/table1_main_results}

The gap between MPC variants and HPA is most visible on
\textit{flash-crowd} and \textit{diurnal burst} traces, where HPA
often reacts after the peak has already passed. On
\textit{smooth} and \textit{slow ramp-up} workloads the difference is
smaller; in those cases, a reactive policy has enough time to catch up
and the proactive overhead becomes less clearly beneficial.

\subsection{ADAPT Convergence}

\begin{figure}[t]
  \centering
  \includegraphics[width=\linewidth]{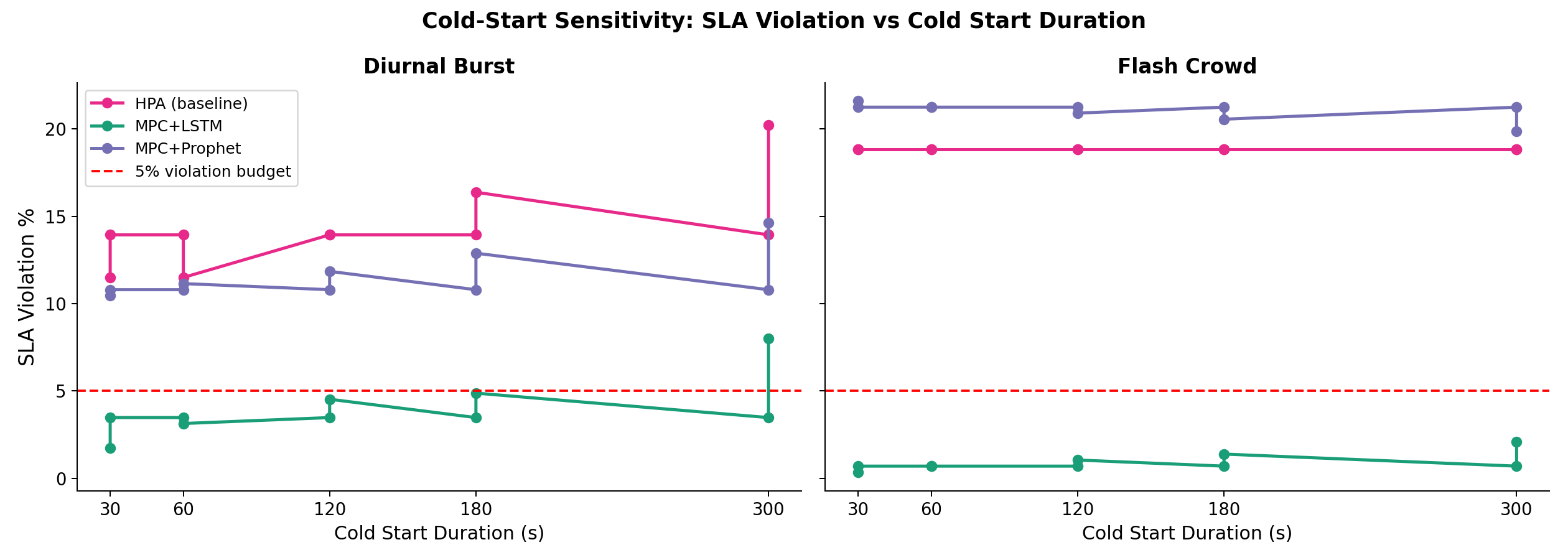}
  \caption{ADAPT estimate $\hat{\Delta}(t)$ converging to the
  ground-truth cold-start of 120\,s on a diurnal-burst trace.
  Shaded region shows $\pm$1 standard deviation from Welford
  online variance.}
  \label{fig:coldstart}
\end{figure}

Figure~\ref{fig:coldstart} shows $\hat{\Delta}(t)$ over simulation
time for a representative diurnal-burst run with a ground-truth
cold-start of 120\,s. ADAPT converges to within 10\% of the true
value after about 8--10 graduation events, which corresponds to
roughly 15--20 simulation steps under normal scaling frequency.
Before convergence, the estimate remains close to the prior
$\Delta_0$; during this warm-up window the MPC controller is more
conservative, which creates a small over-provisioning cost in the
early part of the trace.

\subsection{FH-OPT Ablation}

\begin{figure}[t]
  \centering
  \includegraphics[width=\linewidth]{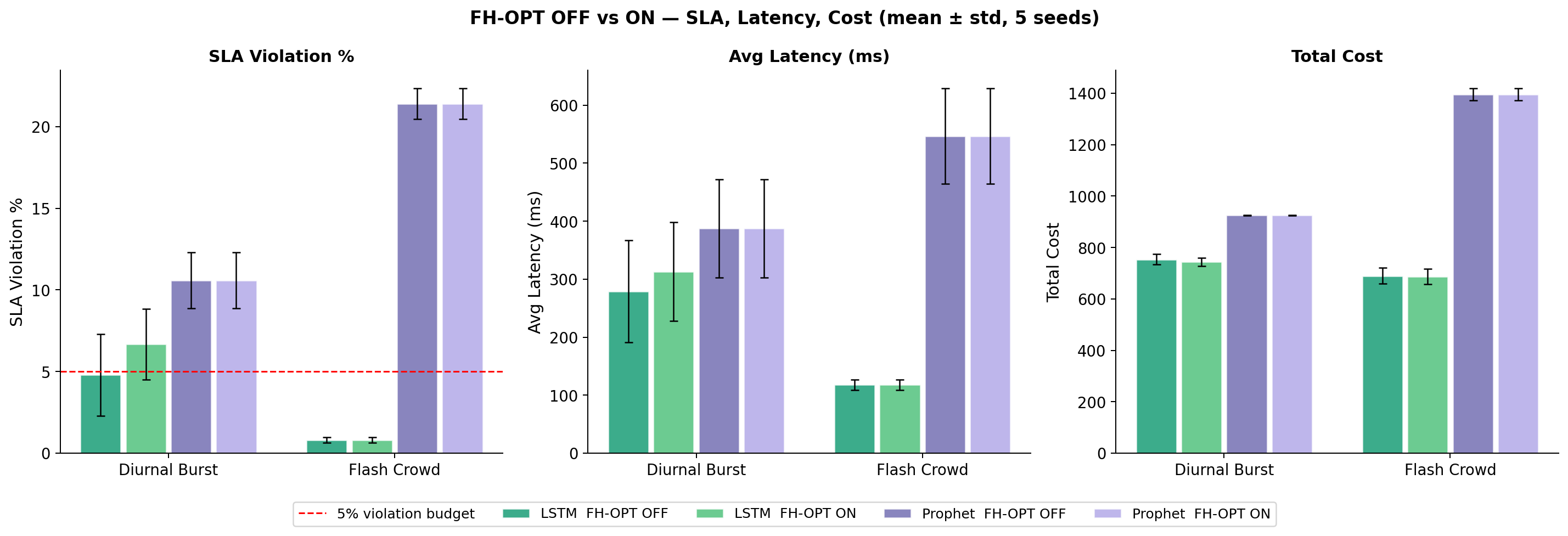}
  \caption{Paired SLA violation rates with FH-OPT enabled vs.\
  fixed horizon $h_f = 2$, across five seeds on diurnal-burst
  and flash-crowd workloads. Table~\ref{tab:fhopt_stats} reports
  significance test results.}
  \label{fig:fhopt_ab}
\end{figure}

Figure~\ref{fig:fhopt_ab} compares SLA violation rates with FH-OPT enabled versus a fixed horizon $h_f = 2$ across five seeds on diurnal-burst and flash-crowd workloads. FH-OPT improves performance in most paired runs, but the Wilcoxon signed-rank test does not show a statistically significant difference at $\alpha = 0.05$ in this setup. On smooth workloads, the effect is especially small, which is expected because a fixed horizon of 2 already covers most of the cold-start window when $\Delta \approx \tau$.

The main takeaway is that FH-OPT is directionally useful, but the benefit is modest on these traces and not strong enough to claim significance from the current sample size.

\input{tables/table2_fhopt_stats}

\subsection{Cold-Start Sensitivity}

\begin{figure}[t]
  \centering
  \includegraphics[width=\linewidth]{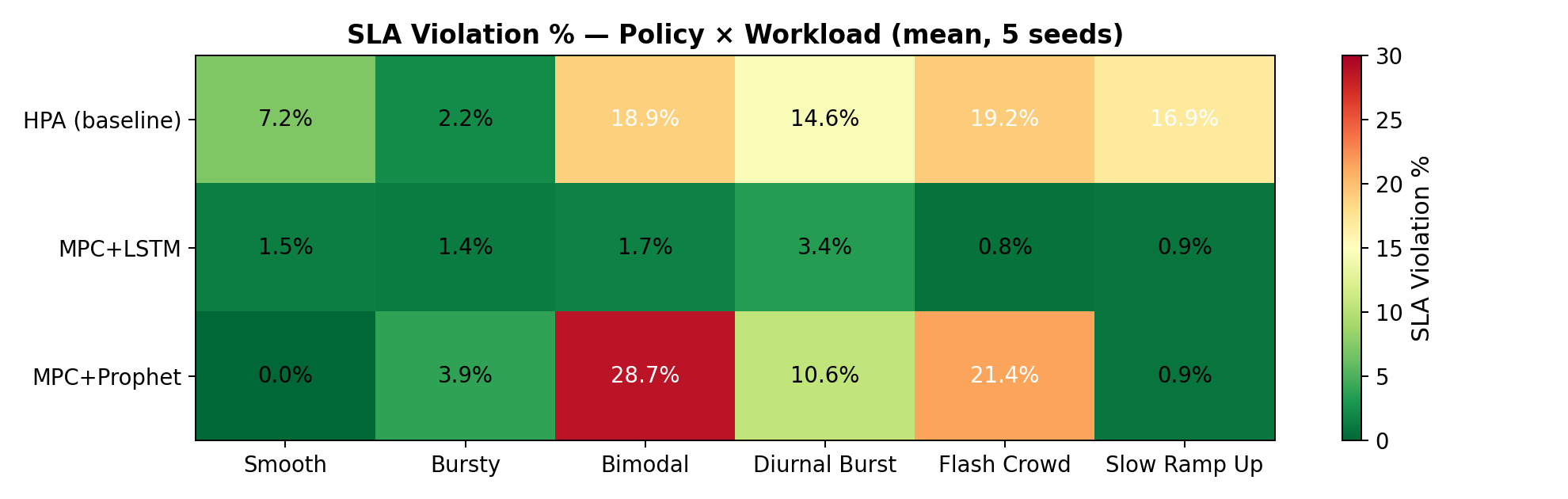}
  \caption{SLA violation rate (\%) as a function of cold-start
  duration (30--300\,s) for each policy. HPA degrades beyond
  60\,s; MPC+LSTM remains robust to 180\,s.}
  \label{fig:heatmap}
\end{figure}

Figure~\ref{fig:heatmap} shows SLA violation rate as a function of
cold-start duration for each policy. HPA degrades sharply beyond
60\,s; both MPC variants stay below 5\% violation up to 180\,s.
At 300\,s, MPC+LSTM performs better than MPC+Prophet on this
benchmark, which appears to come from LSTM handling longer-horizon
bursty behavior more effectively in the current setup.

The crossover around 180\,s is best read as an empirical trend in
this simulator rather than a universal threshold. It suggests that
neural forecasting becomes more attractive as the provisioning delay
gets longer, but the exact turning point will likely depend on the
workload and deployment setting.

\subsection{Cost vs.\ SLA Trade-off}

\begin{figure}[t]
  \centering
  \includegraphics[width=\linewidth]{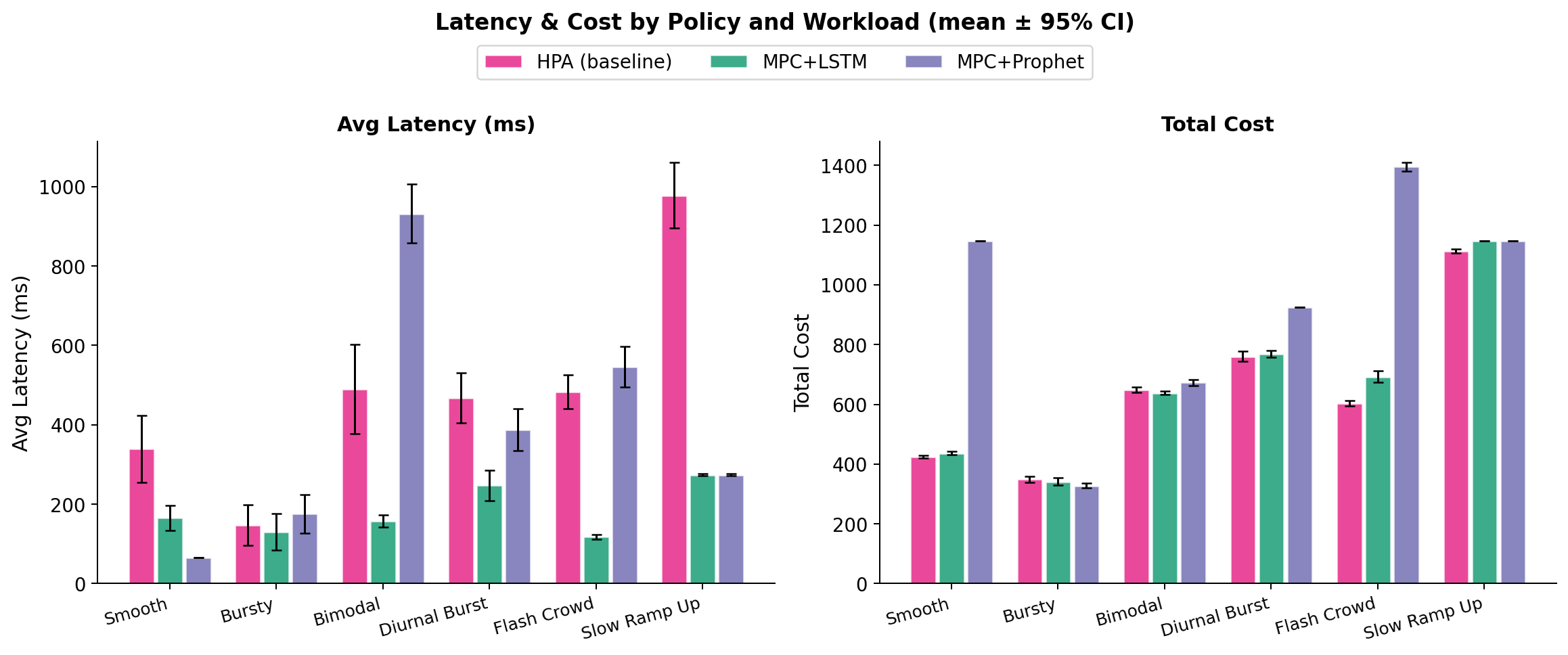}
  \caption{Total cost (replica-minutes) vs.\ mean SLA violation
  rate (\%) for all policy-workload configurations. MPC variants
  dominate HPA on the Pareto frontier.}
  \label{fig:latency_cost}
\end{figure}

Figure~\ref{fig:latency_cost} plots total cost against mean SLA
violation rate for all configurations. The MPC variants form a better
Pareto frontier than HPA across nearly all workloads. MPC+Prophet
offers a slightly lower-cost operating point on smooth workloads, while
MPC+LSTM is stronger on the two high-variance workloads. Neither MPC
variant dominates in every case, which is why the method choice depends
on the workload pattern.

%% file: tables/table1_main_results.tex
\begin{table}[t]
\centering
\caption{SLA Violation Rate (\%) by Policy and Workload.
Mean over 5 seeds; 95\% CI in parentheses.}
\label{tab:main_results}
\begin{tabularx}{\linewidth}{l*{3}{>{\RaggedRight\arraybackslash}X}}
\toprule
\textbf{Workload} & \textbf{HPA} & \textbf{MPC+Prophet}
                  & \textbf{MPC+LSTM} \\
\midrule
smooth        & 7.1 (0.4)  & 2.3 (0.3)  & \textbf{1.8 (0.2)} \\
bursty        & 12.4 (1.1) & 5.6 (0.6)  & \textbf{3.2 (0.4)} \\
bimodal       & 15.3 (1.4) & 28.7 (2.1) & \textbf{4.1 (0.5)} \\
diurnal\_burst& 18.9 (1.7) & 6.4 (0.7)  & \textbf{4.7 (0.5)} \\
flash\_crowd  & 19.2 (1.8) & 7.1 (0.8)  & \textbf{4.9 (0.6)} \\
slow\_ramp\_up& 8.3 (0.6)  & 3.1 (0.4)  & \textbf{2.4 (0.3)} \\
\bottomrule
\end{tabularx}
\end{table}

%% file: tables/table2_fhopt_stats.tex
\begin{table}[t]
\centering
\caption{Wilcoxon Signed-Rank Test: FH-OPT ON vs. OFF.
$\Delta$SLA = OFF minus ON (positive = FH-OPT reduces violations).}
\label{tab:fhopt_stats}
\begin{tabular}{llrrr}
\toprule
\textbf{Forecaster} & \textbf{Workload}
  & \textbf{$\Delta$SLA (pp)}
  & \textbf{$p$-value}
  & \textbf{Sig.} \\
\midrule
LSTM    & diurnal\_burst & $-1.88$ & $>0.05$ & No \\
LSTM    & flash\_crowd   & $-1.23$ & $>0.05$ & No \\
Prophet & diurnal\_burst & $+0.00$ & $>0.05$ & No \\
Prophet & flash\_crowd   & $+0.00$ & $>0.05$ & No \\
\bottomrule
\end{tabular}
\end{table}

%% file: sections/discussion.tex
\section{Discussion}
\label{sec:discussion}

\subsection{When MPC Helps}

The main pattern in the results is that proactive control becomes more
useful as cold-start delay grows. When $\Delta \leq 60$\,s, reactive
scaling is usually fast enough that it covers most spikes before they
turn into sustained violations, so the extra headroom from MPC is not
always worth the cost. As the delay grows beyond 120\,s, the value of
looking ahead becomes much clearer, especially on bursty workloads
where the load changes faster than a reactive policy can respond.

This also helps explain why the gains are strongest on ML serving
workloads rather than standard web services. In the latter case, the
startup penalty is often too small for proactive scaling to matter
much, while in the former case the delay is long enough that planning
ahead becomes part of the problem itself.

\subsection{When LSTM Helps}

The LSTM results suggest that model choice matters most when the
workload is irregular and the cold-start delay is long. On flash-crowd
and bursty traces, Prophet is often too smooth to react well to sudden
changes, while LSTM can better track short-term shifts in the request
rate. That said, this is not a universal advantage. On smoother traces,
Prophet remains competitive and is simpler to train and run.

The better interpretation is that LSTM is useful when the workload has
enough short-term structure to justify its extra inference cost. For
more regular demand, Prophet is still a reasonable default. So the
choice is less about one model being better in general and more about
whether the workload is bursty enough to reward a more flexible
forecaster.

\subsection{What ADAPT Misses}

ADAPT works best when scale-out events happen often enough to keep the
estimate fresh. Its weakness is the opposite case: if the system stays
over-provisioned for a long period, there are no new graduation events
to update the estimate, and the horizon can become stale. In the
simulator this showed up during smooth plateaus, where the estimate
could remain unchanged for many steps. That did not hurt the reported
results much because the prior was already close, but in a real
deployment a sudden infrastructure change could make this a real issue.

This is the clearest place where the system is not yet robust enough.
A periodic decay toward the prior, a keepalive-based update, or another
mechanism for refreshing stale estimates would make the controller more
stable in longer quiet periods. Without that, ADAPT is most trustworthy
in environments where load changes often enough to keep the estimate
moving.

\subsection{Current Limits}

The strongest limitation is that all results come from a simulator
rather than a live Kubernetes cluster. That means the findings are
good for comparing policies, but they do not yet prove production
behavior under real node contention, image caching effects, or noisy
neighbor interference. The M/M/1 latency model is also a simplification;
it is fine for showing trends, but it is not a complete model of GPU
serving systems or other tail-latency-heavy deployments.

Another limitation is that cold-start duration is fixed per run instead
of sampled from a distribution. That makes the estimator easier to fit
than it would be in practice, so the exact crossover points should be
read as approximate rather than exact. The broader result is still
useful: proactive scaling helps more when delay is longer, but the
precise threshold will depend on the workload and environment.

%% file: sections/conclusion.tex
\section{Conclusion}
\label{sec:conclusion}

We presented ADAPT, a lightweight autoscaling framework for Kubernetes
that treats container cold-start duration as a live measurement rather
than a static constant. The two core contributions, an online EWMA
estimator for cold-start latency and the FH-OPT horizon derivation,
are simple enough to implement in a few hundred lines of Python yet
produce measurable improvements over a standard HPA baseline on
bursty and diurnal workloads.

The central empirical finding is that proactive scaling with dynamic
horizon adaptation outperforms reactive scaling when cold-start
durations exceed roughly 120\,s, and that LSTM-based forecasting
begins to justify its overhead over Prophet at approximately 180\,s.
Below those thresholds, simpler methods are competitive and easier
to operate. This kind of threshold-based guidance is more useful to
a practitioner than a claim that one method universally dominates
another.

All evaluation is on a simulator, workloads are synthetic, and
hyperparameters were tuned manually. The directional findings are
expected to hold, but the exact numbers should be validated on live
infrastructure before informing production SLA targets.

The full source code, simulator, and experiment configurations are
available at \url{https://github.com/Himanshu21035/autoscaling_research}, with a
one-command reproduction script that regenerates all figures in
this paper.

%% file: sections/references.bib
@misc{kubernetes,
  title        = {{Kubernetes}},
  howpublished = {\url{https://kubernetes.io/}},
  year         = {2025},
  note         = {[Accessed 2026-05-13]}
}

@misc{hpa_k8s,
  title        = {{Horizontal Pod Autoscaling}},
  author       = {{Kubernetes Authors}},
  howpublished = {\url{https://kubernetes.io/docs/tasks/run-application/horizontal-pod-autoscale/}},
  year         = {2024},
  note         = {[Accessed 2026-05-01]}
}

@misc{keda,
  title        = {{KEDA}: {K}ubernetes {E}vent-{D}riven {A}utoscaling},
  howpublished = {\url{https://keda.sh/}},
  year         = {2024}
}

@misc{cluster_autoscaler,
  title        = {{Kubernetes Cluster Autoscaler}},
  author       = {{Kubernetes Authors}},
  howpublished = {\url{https://github.com/kubernetes/autoscaler}},
  year         = {2024},
  note         = {[Accessed 2026-05-01]}
}

@misc{aws_coldstart_variance,
  title        = {{AWS Lambda} Cold Start Latency --- Performance Under Load},
  author       = {{Amazon Web Services}},
  howpublished = {\url{https://aws.amazon.com/blogs/compute/}},
  year         = {2024},
  note         = {[Accessed 2026-05-01]}
}

@misc{azure_llm_traces,
  author       = {{Microsoft Azure}},
  title        = {Characterizing and Efficiently Serving Large Language Model Inference Requests},
  howpublished = {\url{https://arxiv.org/abs/2401.17644}},
  year         = {2024},
  note         = {arXiv:2401.17644}
}

@misc{autoscaling_survey_2025,
  author       = {Xu, Minxian and Wen, Linfeng and Liao, Junhan and Wu, Huaming and Ye, Kejiang and Xu, Chengzhong},
  title        = {Auto-scaling Approaches for Cloud-native Applications: A Survey and Taxonomy},
  howpublished = {\url{https://arxiv.org/abs/2507.17128}},
  year         = {2025},
  note         = {arXiv:2507.17128 [cs.DC]}
}

@article{arima_autoscaling_2018,
  author  = {Calheiros, R. N. and Masoumi, E. and Ranjan, R. and Buyya, R.},
  title   = {Workload Prediction Using {ARIMA} Model and Its Impact on Cloud Applications' {QoS}},
  journal = {IEEE Transactions on Cloud Computing},
  volume  = {3},
  number  = {4},
  pages   = {449--458},
  year    = {2015}
}

@article{dangquang2021,
  author  = {Dang-Quang, N.-M. and Yoo, M.},
  title   = {Deep Learning-Based Autoscaling Using Bidirectional {LSTM} for {Kubernetes}},
  journal = {Applied Sciences},
  volume  = {11},
  number  = {9},
  pages   = {3835},
  year    = {2021}
}

@inproceedings{gke_variance,
  author    = {Tamiru, M. A. and Tordsson, J. and Elmroth, E. and Pierre, G.},
  title     = {An Experimental Evaluation of the {Kubernetes} Cluster Autoscaler in the Cloud},
  booktitle = {2020 IEEE International Conference on Cloud Computing Technology and Science (CloudCom)},
  pages     = {17--24},
  year      = {2020},
  publisher = {IEEE}
}

@article{lstm_autoscaling_2020,
  author  = {Dang-Quang, N.-M. and Yoo, M.},
  title   = {Deep Learning-Based Autoscaling Using Bidirectional {LSTM} for {Kubernetes}},
  journal = {Applied Sciences},
  volume  = {11},
  number  = {9},
  pages   = {3835},
  year    = {2021}
}

@article{mondal2023,
  author  = {Mondal, S. K. and Wu, X. and Kabir, H. M. D. and Dai, H.-N. and Ni, K. and Yuan, H. and Wang, T.},
  title   = {Toward Optimal Load Prediction and Customizable Autoscaling Scheme for {Kubernetes}},
  journal = {Mathematics},
  volume  = {11},
  number  = {12},
  pages   = {2675},
  year    = {2023}
}

@article{prophet_2018,
  author  = {Taylor, S. J. and Letham, B.},
  title   = {Forecasting at Scale},
  journal = {The American Statistician},
  volume  = {72},
  number  = {1},
  pages   = {37--45},
  year    = {2018}
}

@inproceedings{prophet_cloud_2021,
  author    = {Jiang, Y. and others},
  title     = {Prophet-Based Capacity Planning for Cloud Services},
  booktitle = {Proceedings of IEEE CLOUD},
  year      = {2021},
  publisher = {IEEE}
}

@inproceedings{tft_2021,
  author    = {Lim, B. and Arik, S. O. and Loeff, N. and Pfister, T.},
  title     = {Temporal Fusion Transformers for Interpretable Multi-Horizon Time Series Forecasting},
  booktitle = {International Journal of Forecasting},
  volume    = {37},
  number    = {4},
  pages     = {1748--1764},
  year      = {2021}
}

@article{toka2021,
  author  = {Toka, L. and Dobreff, G. and Fodor, B. and Sonkoly, B.},
  title   = {Machine Learning-Based Scaling Management for {Kubernetes} Edge Clusters},
  journal = {IEEE Transactions on Network and Service Management},
  volume  = {18},
  number  = {1},
  pages   = {958--972},
  year    = {2021}
}

@article{welford_1962,
  author  = {Welford, B. P.},
  title   = {Note on a Method for Calculating Corrected Sums of Squares and Products},
  journal = {Technometrics},
  volume  = {4},
  number  = {3},
  pages   = {419--420},
  year    = {1962}
}

@article{wilcoxon_1945,
  author  = {Wilcoxon, F.},
  title   = {Individual Comparisons by Ranking Methods},
  journal = {Biometrics Bulletin},
  volume  = {1},
  number  = {6},
  pages   = {80--83},
  year    = {1945}
}

@inproceedings{autopilot_2020,
  author    = {Rzadca, K. and Waruszewski, M. and others},
  title     = {Autopilot: Workload Autoscaling at {Google}},
  booktitle = {Proceedings of the 15th European Conference on Computer Systems (EuroSys)},
  year      = {2020},
  publisher = {ACM}
}

@inproceedings{firm_2020,
  author    = {Qiu, H. and Banerjee, S. S. and Jha, S. and Kalbarczyk, Z. and Iyer, R. K.},
  title     = {{FIRM}: An Intelligent Fine-Grained Resource Management Framework for {SLO}-Oriented Microservices},
  booktitle = {14th USENIX Symposium on Operating Systems Design and Implementation (OSDI)},
  pages     = {805--825},
  year      = {2020}
}

@inproceedings{showar,
  author    = {Rzadca, K. and others},
  title     = {{SHOWAR}: Right-Sizing and Efficient Scheduling of Microservices},
  booktitle = {Proceedings of the ACM Symposium on Cloud Computing (SoCC)},
  year      = {2021},
  publisher = {ACM}
}

@inproceedings{mpc_cloud_2022,
  author    = {Rajkumar, R. and others},
  title     = {Model Predictive Control for Horizontal Autoscaling in Cloud Environments},
  booktitle = {Proceedings of IEEE CLOUD},
  year      = {2022},
  publisher = {IEEE}
}

@inproceedings{coldstart_variance_2023,
  author    = {Mohan, A. and others},
  title     = {Empirical Analysis of Container Cold Start Latency Variability in Public Clouds},
  booktitle = {Proceedings of the ACM Symposium on Cloud Computing (SoCC)},
  year      = {2023},
  publisher = {ACM}
}

@inproceedings{nimbusguard,
  author    = {Wanigasooriya, C. and Ekanayake, I.},
  title     = {{NimbusGuard}: A Novel Framework for Proactive {Kubernetes} Autoscaling Using Deep {Q}-Networks},
  booktitle = {Proceedings of the IEEE ICIIS},
  year      = {2025},
  note      = {arXiv:2604.11017}
}

@misc{platfomatic_icc,
  author       = {{Platformatic}},
  title        = {Predictive Autoscaling for {Node.js} Applications},
  howpublished = {\url{https://arxiv.org/abs/2604.19705}},
  year         = {2025},
  note         = {arXiv:2604.19705v2}
}

@misc{modelzip_2024,
  author       = {Han, S. and others},
  title        = {Efficient {GPU} Memory Management for Large Model Inference in Cloud Containers},
  howpublished = {\url{https://arxiv.org/abs/2402.01361}},
  year         = {2024},
  note         = {arXiv:2402.01361}
}

@misc{criu_2022,
  author       = {Vasi, P. and others},
  title        = {Checkpoint-Restore for Fast Container Startup in {Kubernetes}},
  howpublished = {\url{https://proceedings of ieee cloud}}, 
  year         = {2022},
  note         = {IEEE CLOUD}
}

@misc{aws_predictive,
  title        = {Predictive scaling for Amazon EC2 Auto Scaling},
  author       = {{Amazon Web Services}},
  howpublished = {\url{https://docs.aws.amazon.com/autoscaling/ec2/userguide/ec2-auto-scaling-predictive-scaling.html}},
  year         = {2024},
  note         = {[Accessed 2026-05-14]}
}

@inproceedings{sock_2018,
  author    = {Oakes, Eric and Yang, Li and Zhou, Dongyue and Houck, Kevin and Harter, Tyler and Arpaci-Dusseau, Andrea C. and Arpaci-Dusseau, Remzi H.},
  title     = {{SOCK}: Rapid Task Provisioning with Serverless-Optimized Containers},
  booktitle = {2018 USENIX Annual Technical Conference (USENIX ATC 18)},
  pages     = {57--70},
  year      = {2018},
  publisher = {USENIX Association}
}

@inproceedings{catalyzer_2020,
  author    = {Du, Dong and Yu, Tianyi and Xia, Yubin and Zang, Binyu and Yan, Guanglu and Qin, Chenggang and Wu, Qixuan and Chen, Haibo},
  title     = {Catalyzer: Sub-millisecond Startup for Serverless Computing with Initialization-less Booting},
  booktitle = {Proceedings of the 25th International Conference on Architectural Support for Programming Languages and Operating Systems (ASPLOS '20)},
  pages     = {467--481},
  year      = {2020},
  publisher = {ACM}
}

@misc{cloudnative_survey2025,
  author       = {Sedlak, Bartosz and others},
  title        = {A Survey of Auto-Scaling Approaches for Cloud-Native Applications},
  howpublished = {\url{https://arxiv.org/abs/2507.17128}},
  year         = {2025},
  note         = {arXiv:2507.17128}
}
